\pdfoutput=1

\documentclass[11pt]{article}

\usepackage[final]{acl}

\usepackage{times}
\usepackage{latexsym}
\usepackage{amsmath} 
\usepackage[T1]{fontenc}

\usepackage[utf8]{inputenc}

\usepackage{microtype}

\usepackage{inconsolata}

\usepackage{graphicx}

\title{Synergizing Logical Reasoning, Knowledge Management and Collaboration in Multi-Agent LLM System}

\author{
  \textbf{Adam Kostka} \and
  \textbf{Jarosław A. Chudziak}
  \\
  Faculty of Electronics and Information Technology\\
  Warsaw University of Technology, Poland \\
  \texttt{adam.kostka.stud@pw.edu.pl} \and \texttt{jaroslaw.chudziak@pw.edu.pl}
}

\begin{document}
\maketitle
\begin{abstract}
This paper explores the integration of advanced Multi-Agent Systems (MAS) techniques to develop a team of agents with enhanced logical reasoning, long-term knowledge retention, and Theory of Mind (ToM) capabilities. By uniting these core components with optimized communication protocols, we create a novel framework called SynergyMAS, which fosters collaborative teamwork and superior problem-solving skills. The system’s effectiveness is demonstrated through a product development team case study, where our approach significantly enhances performance and adaptability. These findings highlight SynergyMAS's potential to tackle complex, real-world challenges.
\end{abstract}


\section{Introduction}

Large Language Models (LLMs) have seen significant advancements in recent years, particularly in answer accuracy, increased context windows, and the ability to tackle complex tasks. However, despite these improvements, LLMs continue to face challenges such as hallucinations, knowledge gaps due to incomplete training data, and difficulties in managing long-term dependencies \citep{naveed2024comprehensiveoverviewlargelanguage}. Enhancing LLM performance remains complex, often requiring enormous resources for training and domain-specific data integration, which may not always yield the desired outcomes \citep{Raschka_2024}.

A promising approach to overcoming these limitations is the development of Multi-Agent Systems (MAS) that incorporate LLMs at their core. MAS can be tailored for specific use cases, allowing more sophisticated solutions by refining communication protocols and integrating external tools and databases.

This study aims to broaden the capabilities of LLMs by creating a MAS framework that integrates three critical components: logical reasoning, Retrieval-Augmented Generation (RAG), and Theory of Mind (ToM). By combining these elements, we propose a new framework named SynergyMAS, which enhances LLM capabilities for complex tasks. The framework was tested on multiple LLMs, including Claude and Gemini, to evaluate its versatility and effectiveness.

\begin{figure}[t]
\centerline{\includegraphics[width=0.5\textwidth]{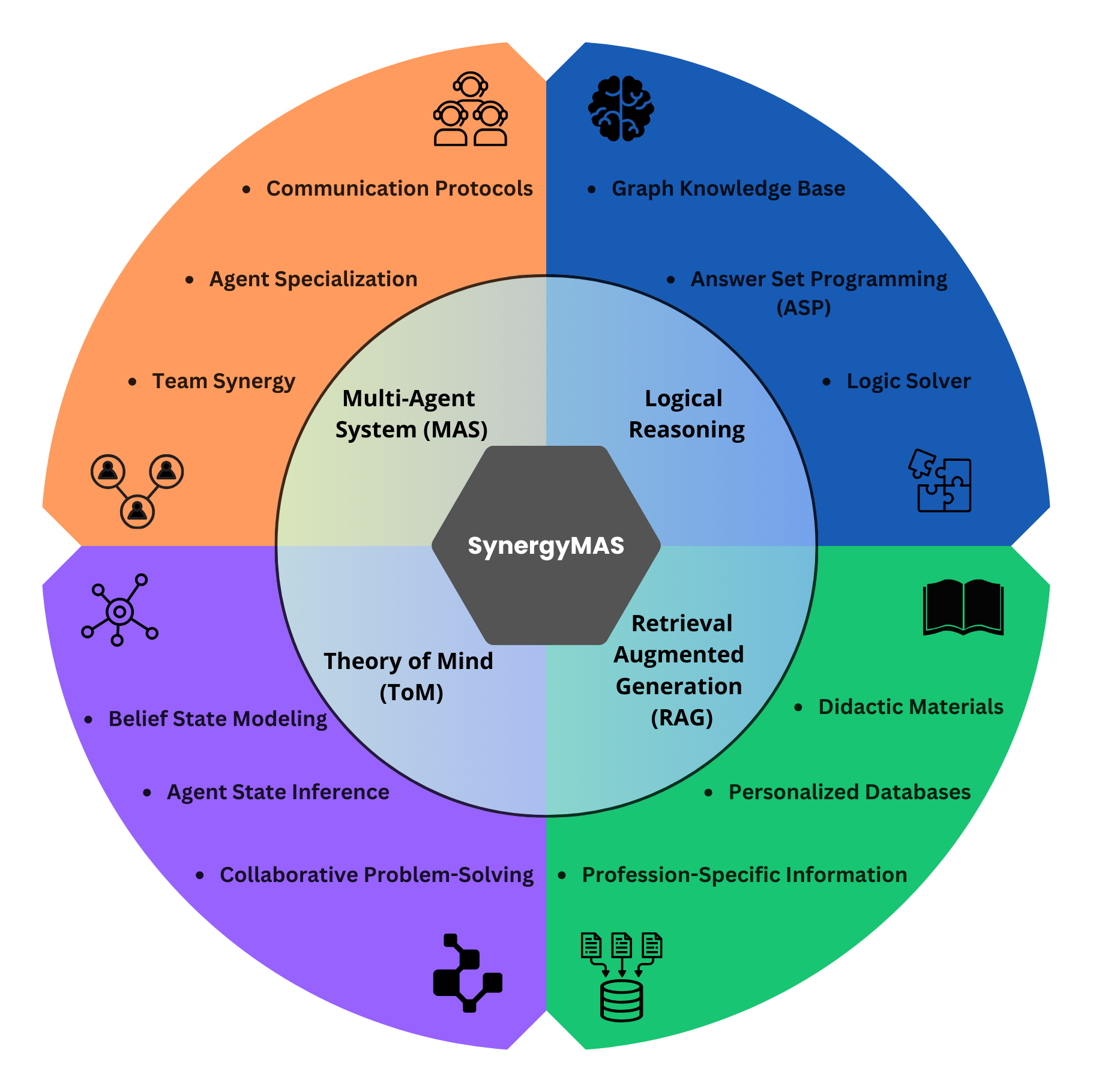}}
\caption{SynergyMAS: Integrating Logical Reasoning, RAG, and Theory of Mind in a Multi-Agent System to enhance LLM capabilities for complex tasks (based on \citet{sun2024surveyreasoningfoundationmodels}).}
\label{fig:pie_diagram}
\end{figure}

LLMs often struggle with complex reasoning tasks that require logical thinking, frequently leading to hallucinations \citep{chen2023learningteachlargelanguage}. Previous research has shown that incorporating logical reasoning, such as a logical solver, can mitigate these issues on logic testing datasets \citep{pan-etal-2023-logic}. In this study, we implement a graph knowledge base integrated with a logic solver using Answer Set Programming (ASP), enhancing the system's ability to engage in extended, logical discussions.

Moreover, MAS members will have access to a personalized RAG database with didactic materials specific to their profession. This approach aims to improve specialists' performance by providing them with relevant information without overburdening others.

As LLMs evolve, their ToM capabilities also improve \citep{kosinski2024evaluatinglargelanguagemodels}. Given the importance of ToM in collaborative problem-solving, enhancing these capabilities should boost system effectiveness. Including a belief state in agents' communications allows for better inference of each other's states, promoting better teamwork.

This paper contributes to Artificial Intelligence, Machine Learning, and Dialogue Systems by:
\begin{enumerate}
    \item Proposing a novel architecture for multi-agent LLM systems, integrating logical solvers with graph databases and new communication protocols that enhance agent synergy.
    \item Demonstrating the effectiveness of combining logical reasoning, RAG, and ToM in improving agent capabilities for complex conversations requiring external knowledge and domain-specific expertise.
    \item Offering insights into practical applications of such systems in collaborative problem-solving scenarios, identifying real-world contexts where SynergyMAS can significantly enhance system performance.
\end{enumerate}

The main components of SynergyMAS are illustrated in Figure \ref{fig:pie_diagram}.

\section{Related Work}

SynergyMAS integrates several ideas discussed in various studies, each playing a critical role in enhancing MAS performance. This section explores the current research on these components.

\subsection{Background on Multi-Agent LLM Systems}
\begin{figure}[t]
\centerline{\includegraphics[width=0.25\textwidth]{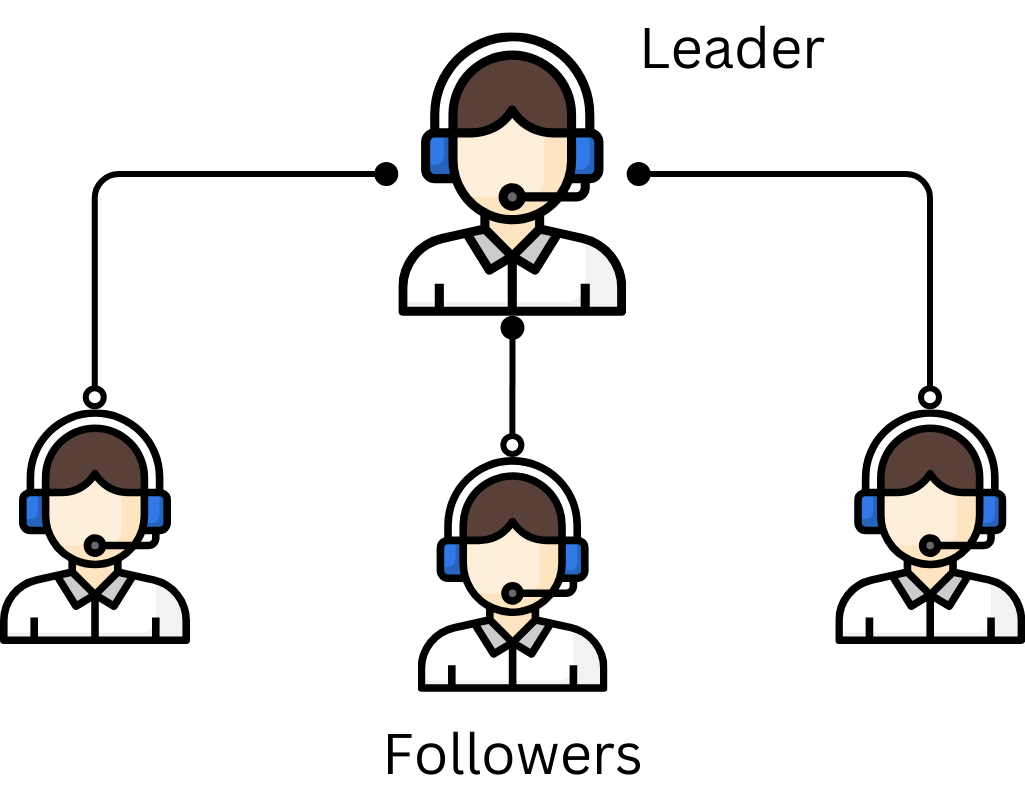}}
\caption{Hierarchical team structure.}
\label{fig:hier}
\end{figure}

The topic of MAS has been ongoing for many years, with significant contributions documented \citep{Wooldridge_2009}. Recently, MAS has gained popularity in AI due to its superior ability to solve complex problems through the collaborative efforts of autonomous agents \citep{lei2024macmutilizingmultiagentcondition}. MAS architectures typically involve a network of intelligent agents with specialized knowledge working towards common objectives \cite{Russell_Norvig_2016}. A hierarchical structure, as showcased in Figure \ref{fig:hier}, is one such structure utilized in this article. Recent improvements in LLMs continue to enhance the potential of MAS, allowing for more refined communication and reasoning among agents \citep{wu2023autogenenablingnextgenllm}. However, as these systems grow in complexity, they face coordination, scalability, and consistency challenges, driving ongoing research to improve MAS functionality.

\subsection{Overview of Logical Reasoning in AI}
The complex and opaque nature of LLMs makes it challenging to directly analyze and improve their behavior and outputs \citep{Calegari_Ciatto_Mascardi_Omicini_2020}. This has made it difficult for LLMs to effectively handle convoluted logical reasoning tasks \citep{chen2024improvinglargelanguagemodels}. Incorporating logical solver tools into MAS can address this limitation \citep{chen2024languagemodelspretendsolvers}. Many studies have attempted to integrate logical thinking into models using single or multiple methods \citep{yang2023couplinglargelanguagemodels}. The typical approach involves converting natural language into a logical format \citep{Gelfond_Kahl_2014}, followed by applying a logic solver to resolve the query. This approach has shown promising results, improving performance on logic testing datasets like ProofWriter \citep{tafjord2021proofwritergeneratingimplicationsproofs}. However, the use of these solvers to enhance reasoning capabilities in multi-agent systems still requires further exploration.

\subsection{Graph Databases in AI Systems}
Graph databases have gained prominence in AI for their ability to efficiently manage and query complex relationships between data points. Unlike traditional databases, graph databases use nodes and edges to represent and traverse connections, making them well-suited for interconnected data applications \citep{Lane_Hapke_Howard_2019}. Graph databases like neo4j are increasingly used in AI to enhance reasoning and decision-making by providing structured, context-aware data retrieval \citep{besta2023demystifyinggraphdatabasesanalysis}. Their integration into AI systems supports advanced tasks such as knowledge management \citep{liang2024aligninglargelanguagemodels} and logical inference, essential for tackling complex problems.

\subsection{RAG (Retrieval-Augmented Generation)}
RAG has emerged as a critical enhancement in the LLM landscape, addressing challenges related to context expansion and the inclusion of real-world knowledge in responses while reducing hallucinations \citep{lewis2021retrievalaugmentedgenerationknowledgeintensivenlp}. This approach allows models to analyze vast amounts of information beyond their training data, leading to the creation of specialized agents with domain-specific expertise. Recent innovations like Corrective RAG (CRAG) \citep{yan2024correctiveretrievalaugmentedgeneration} and GraphRAG \citep{edge2024localglobalgraphrag} demonstrate the field's dynamic nature and progress in implementing long-term knowledge into models. In our study, we employ CRAG, which includes a web search tool that provides agents with relevant real-time data. This synergy expands the agents' knowledge base and improves their expertise. By introducing agents to trade knowledge, we aim to closely mirror the behavior and insights of real-world specialists.

\subsection{Theory of Mind}
ToM enables agents to reason about others' mental states \citep{McLaughlin_Beckermann_Walter_2011}, facilitating effective collaboration and communication. In MAS, ToM allows agents to infer others' beliefs, intentions, and goals \citep{Shoham_Leyton-Brown_2012}. While LLMs have shown promising results in ToM tasks \citep{Li_2023}, further tests are needed in text-based problem-solving scenarios. ToM is crucial for agent collaboration, enabling them to understand each other's perspectives and make collective decisions. By incorporating ToM capabilities into our system, we aim to improve agents' social intelligence, making them more effective collaborators.

\section{System Architecture: SynergyMAS}

\begin{figure}[t]
\centerline{\includegraphics[width=0.5\textwidth]{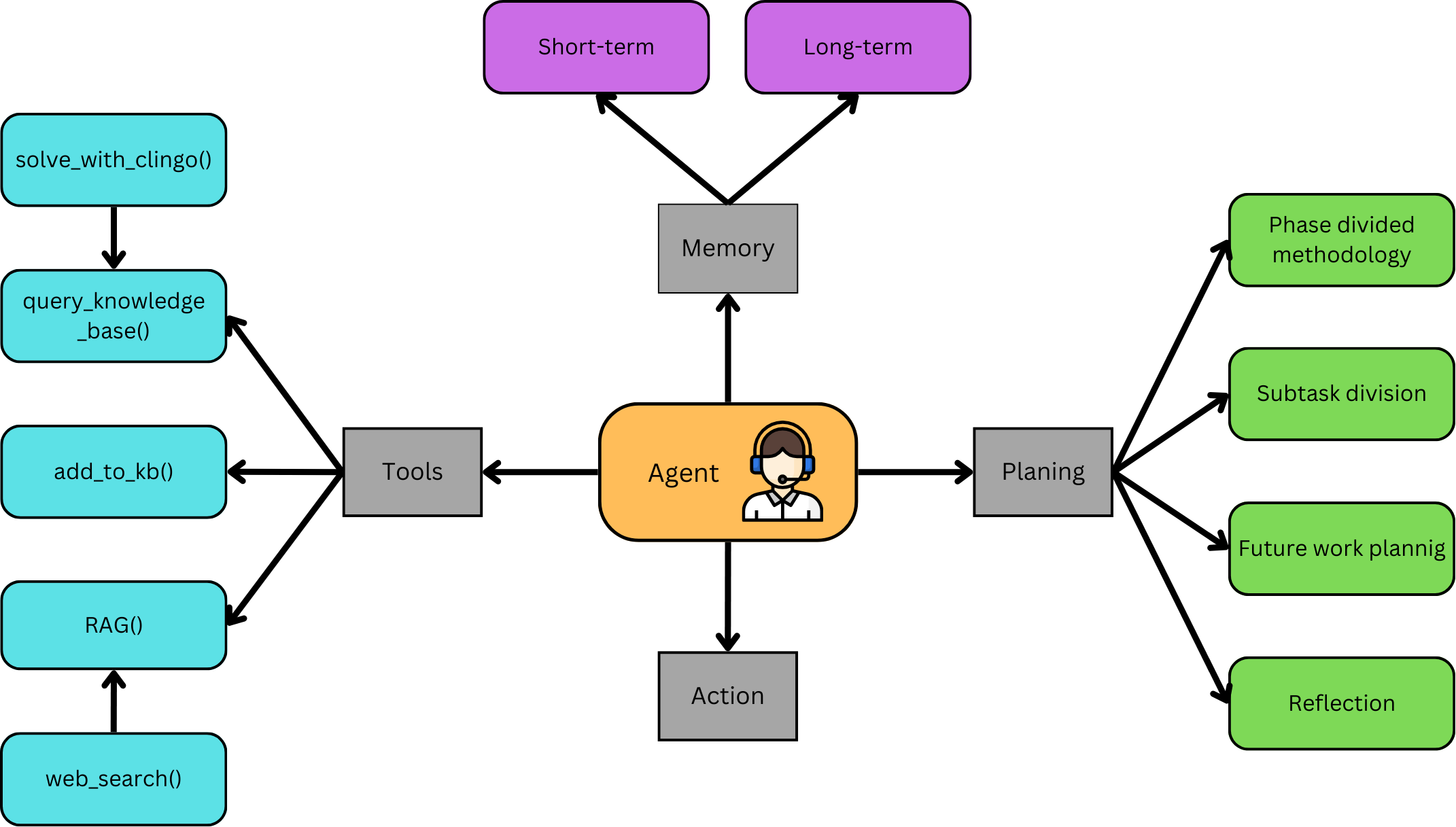}}
\caption{Agents architecture: illustrates the SynergyMAS architecture, showcasing the key components and interactions of an agent within the multi-agent system (based on \citet{weng2023agent}).}
\label{fig:agent_architecture}
\end{figure}

SynergyMAS is designed to leverage multiple agents working together to solve complex, domain-specific problems. Figure \ref{fig:agent_architecture} illustrates the architecture of a single agent within this system, highlighting key components such as memory management, planning capabilities, knowledge retrieval and reasoning tools, and action execution. This agent structure forms the foundation of SynergyMAS and underpins the functionality discussed in the following subsections.

\subsection{System Philosophy and Objectives}

The development of SynergyMAS is guided by a central question: How can a multi-agent system powered by Large Language Models (LLMs) effectively navigate and solve complex, domain-specific problems? The solution is to build a system capable of managing long, detailed conversations while addressing the unique characteristics of specialized domains. 

To achieve this, SynergyMAS was designed with its core components of logical reasoning, knowledge retrieval, and Theory of Mind working together in synergy. This synergy is also about employing communication protocols crafted to unify individual agents' efforts, amplifying each approach's strengths. The system's hierarchical structure, with a central "boss" agent coordinating tasks, ensures all agents operate together in a constructive fashion. 

The success of SynergyMAS is determined by its ability to maintain logical consistency, efficiently retrieve and synthesize knowledge, and adapt dynamically to evolving problem-solving scenarios. The true synergy in SynergyMAS lies in this orchestrated union, where each element enhances the others, resulting in a system with superior capabilities.

\subsection{Overall System Design}

The overall design of SynergyMAS revolves around a hierarchical structure, with a "boss" agent coordinating the workflow and ensuring that all agents align with the system's objectives. This structure integrates three key components: Logical Reasoning, RAG-Based Knowledge Management, and ToM Capabilities.

\textbf{Logical Reasoning} component is powered by a Neo4j graph knowledge base and a logic solver. Conversation data is continuously added to the graph, enabling agents to retrieve relevant information, which is then translated into an ASP query. The logic solver processes this query and returns a logical response in natural language.

\textbf{RAG-Based Knowledge Management} utilizes a modified version of Corrective RAG (CRAG). This component first retrieves information from a Chroma vector base containing domain-specific data. The query is forwarded to an external web search using the Tavily Search framework if relevant data is not found.

\textbf{ToM capabilities} capabilities are integrated through the "My Beliefs" section in each agent's response. This feature enables agents to infer and reason about the beliefs and strategies of others, promoting effective collaboration and coordination.

\subsection{Agent Interactions}

Agent interactions in SynergyMAS are governed by structured communication protocols, ensuring efficient problem-solving. After each task, control returns to the "boss" agent, who evaluates progress, tracks the conversation stage, and assigns new tasks. This process prevents redundancy and maintains focus on the current objectives.

Each agent's response is divided into three parts: My Beliefs, Response, and Future Work. 

\begin{itemize}
\item 
\textbf{My Beliefs} reflects the agent's understanding of the task, incorporating insights from other agents and applying ToM to align with the team’s overall strategy.

\item 
\textbf{Response} section delivers task-specific solutions, leveraging logical reasoning and RAG for accuracy and relevance.

\item
\textbf{Future Work} outlines potential next steps and challenges, guiding the boss agent in steering the conversation forward without unnecessary deviations.
\end{itemize}
This structured approach ensures clarity, cohesion, and focus throughout the problem-solving process.

\section{Logical Reasoning Component}

\begin{figure}[t]
\centerline{\includegraphics[width=0.35\textwidth]{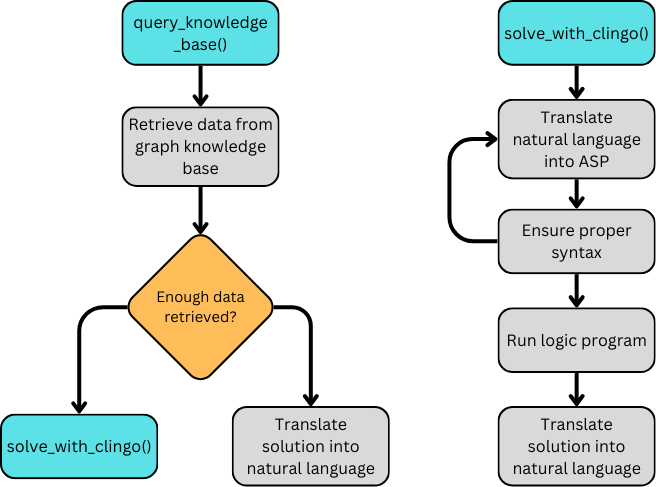}}
\caption{Logical Reasoning Functions: Shows the core logical functions of SynergyMAS, including knowledge base expansion and query solving.}
\label{fig:logic_solver_architecture}
\end{figure}

The logical reasoning component in SynergyMAS integrates a Neo4j graph knowledge base with a logic solver, specifically Clingo, to manage and process complex queries. A logic solver is a tool that processes queries based on formal logic, allowing the system to infer conclusions from a set of given facts and rules. Clingo, a widely-used logic solver, leverages ASP to solve complex problems by generating potential solutions that satisfy all the logical constraints \citep{wang2024chatlogicintegratinglogicprogramming}, as illustrated in Figure \ref{fig:logic_solver_architecture}.

\subsection{Adding and Retrieving Data from the Knowledge Base}

After each agent's response, the \texttt{add\_to\_kb} function is called to process the agent's answer and add relevant data to the Neo4j graph knowledge base.

When an agent needs to retrieve data, the \texttt{query\_knowledge\_base} function is invoked with the agent's question. The question is first translated into a Cypher query by an LLM, which retrieves the relevant data from the graph. If the amount of retrieved data is below a predefined threshold, the query is further translated into an ASP format and processed by the Clingo solver. If the retrieved data exceeds the threshold, it is directly translated into natural language and returned to the agent.

\subsection{Implementation Using Logic Solver}

When data retrieval from the graph is insufficient, the LLM translates the natural language question into a logical representation suitable for processing by the Clingo solver. For example, consider a scenario where the knowledge base includes facts about software development tasks:

\[
\text{task}(\text{ImplementFeatureX})
\]
\[
\forall x (\text{task}(x) \wedge \text{assigned}(x, \text{Alice}) \rightarrow \text{completed}(x))
\]
\[
\text{assigned}(\text{ImplementFeatureX}, \text{Alice})
\]

The query might be: "Is the task ImplementFeatureX completed?" The logic solver would process this and infer:

\[
\text{completed}(\text{ImplementFeatureX})
\]

Safety mechanisms validate the ASP syntax to ensure logical consistency and correctness, allowing the LLM up to three attempts to produce valid code \citep{McGinness}. During testing, the GPT-4o model demonstrated superior performance compared to the GPT-3.5-Turbo model, making it the preferred choice.

After processing by the Clingo solver, the ASP output is translated back into natural language, providing the agent with a clear and accurate response.

\section{RAG Knowledge Base}

\begin{figure}[t]
\centerline{\includegraphics[width=0.25\textwidth]{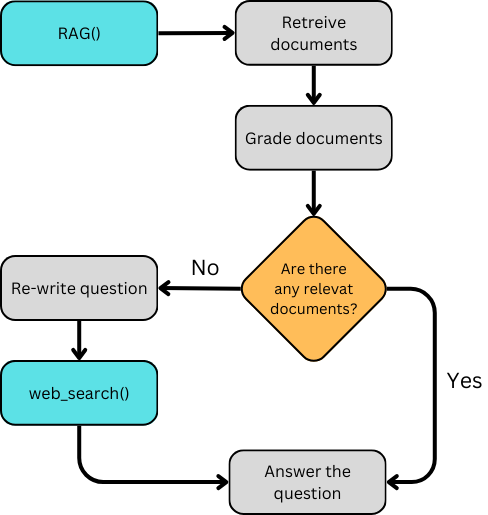}}
\caption{Corrective-RAG flowchart: Document retrieval, grading, and adaptive question-answering process.}
\label{fig:CRAG_architecture}
\end{figure}

The RAG knowledge base in SynergyMAS is designed to provide agents with access to domain-specific knowledge, making their knowledge and actions more similar to those of a domain specialist. 
The system leverages a modified version of CRAG, which enhances standard RAG by incorporating a document grading layer. This layer classifies the relevance of documents stored in a Chroma vector base, ensuring that the most relevant information is retrieved. If the internal database lacks relevant data, the query is forwarded to an external search using the Tavily web search tool.

Figure \ref{fig:CRAG_architecture} illustrates the workflow of the modified CRAG system used in SynergyMAS. The retrieval process begins with query analysis, where the agent formulates a query based on the assigned task. CRAG then grades the relevance of documents within the database, retrieving information from the Chroma vector base if relevant data is available. The query is forwarded to the Tavily Search tool for external data if necessary. The query is optimized for internet search, and retrieved information is synthesized to generate a coherent and accurate response.

The vector-based storage system in SynergyMAS enhances efficiency by enabling quick retrieval and reuse of processed data, allowing integration of documents. This scalable framework, supported by CRAG, ensures agents can access and utilize up-to-date, domain-specific knowledge. By combining internal databases with external web search capabilities, SynergyMAS provides a robust and flexible environment for knowledge retrieval, empowering agents to perform their specialized roles effectively and improving overall system performance.

\section{Theory of Mind Implementation}

In the SynergyMAS system, ToM capabilities enhance collaborative efficiency by enabling agents to reason about their own and others' mental states. ToM is implemented through explicit belief state representations, allowing agents to attribute beliefs, intents, and knowledge to themselves and their teammates.

\begin{figure}[t]
\centerline{\includegraphics[width=0.38\textwidth]{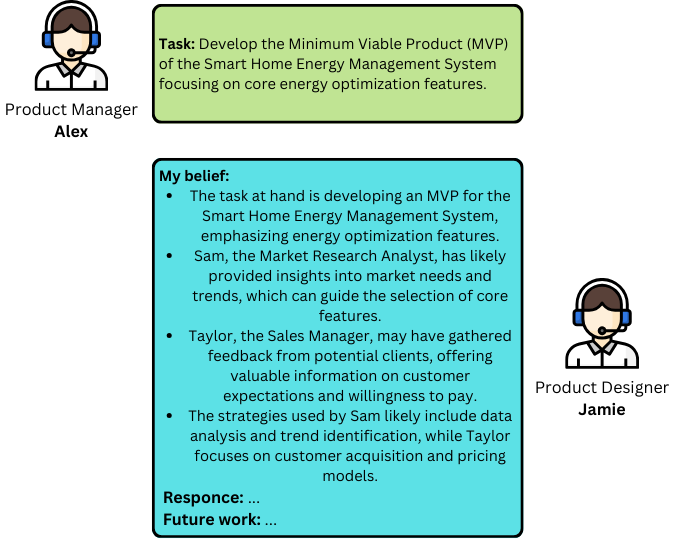}}
\caption{Belief Prompts for Collaboration: Demonstrates how 'My Beliefs' sections facilitate agent collaboration and informed decision-making.}
\label{fig:lsm}
\end{figure}

Inspired by recent studies, the belief state system in SynergyMAS uses prompt engineering to represent and update belief states dynamically. Each agent's belief state evolves with new information and interactions, facilitating a deeper understanding of team dynamics and improving task execution.
Figure \ref{fig:lsm} demonstrates how the 'My Beliefs' sections enable agents to integrate knowledge from different team members, fostering informed decision-making.

ToM capabilities are integrated into the \textit{My Beliefs} section of agent responses, which builds upon two ideas: \begin{itemize} \item \textbf{Introspection:} Agents summarize their own mental states and recent actions, enhancing self-awareness. \item \textbf{First-Order ToM:} Agents infer the mental states of other agents based on shared information and observations.  \end{itemize}

By incorporating ToM, SynergyMAS significantly improves collaborative efficiency and overall system performance. Explicit belief state representations allow for aligned strategies and coherent problem-solving, resulting in better decision-making, more transparent communication, reduced errors, and increased efficiency. This ToM-enhanced approach demonstrates the potential for creating more sophisticated, human-like AI systems that advance multi-agent collaboration across various domains.

\section{Product Development Team Case Study (Lean Startup Methodology)}

\begin{figure}[t]
\centerline{\includegraphics[width=0.5\textwidth]{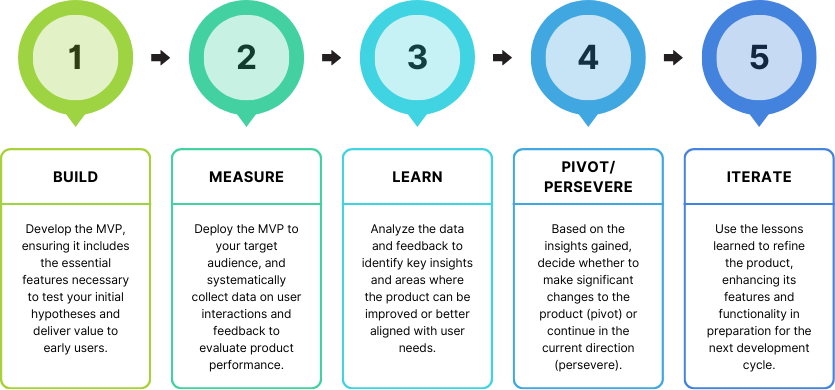}}
\caption{Diagram illustrating the Lean Startup methodology \citep{Ries_2017} steps used in our team case study.}
\label{fig:lean-startup}
\end{figure}

This case study demonstrates SynergyMAS's application in product development using the Lean Startup methodology. While this example focuses on a specific approach, the SynergyMAS framework is versatile and applicable to various problem-solving scenarios across different domains. Figure \ref{fig:lean-startup} illustrates the Lean Startup process used in this case study, highlighting the iterative Build-Measure-Learn cycles that guide product development.

\subsection{Defining the Case Study}

\begin{figure}[t]
\centerline{\includegraphics[width=0.4\textwidth]{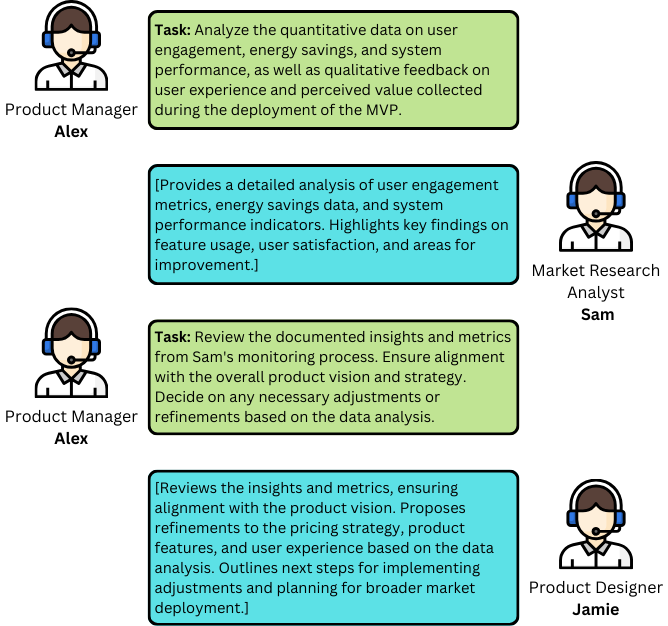}}
\caption{Multi-Agent Task Progression: Illustrates the sequential collaboration among agents to advance the Smart Home Energy Management System from concept to deployment.}
\label{fig:interaction}
\end{figure}

This case study explores how SynergyMAS can be used to develop a Smart Home Energy Management System, employing the Lean Startup methodology \citep{Ries_2017}. The goal is to show how a team of specialized agents can collaboratively guide product development through iterative Build-Measure-Learn cycles. The primary objectives are to validate product-market fit, optimize energy management algorithms, and ensure the system meets market demands and regulatory standards. The sequential collaboration between Alex, Jamie, and Sam is depicted in Figure \ref{fig:interaction}.

\subsection{Team Structure and Roles}

The product development team comprises a Product Manager (PM), a Market Research Analyst (MRA), a Product Designer (PD), and a Sales Manager (SM). Each agent specializes in a distinct domain, contributing their expertise to product development and refinement.

\vspace{0.3cm}

\noindent\textbf{Product Manager (PM) - Alex}

\noindent\textit{Role:} Oversees the Lean Startup process, making key decisions about product direction and ensuring alignment with company goals.

\noindent\textit{Responsibilities:} Alex defines the product vision and strategy, coordinates the Build-Measure-Learn cycles, synthesizes feedback from other agents, and ensures the project stays on track.

\vspace{0.2cm}

\noindent\textbf{Market Research Analyst (MRA) - Sam}

\noindent\textit{Role:} Conducts market analysis and identifies customer needs and trends to inform the product development process.

\noindent\textit{Responsibilities:} Sam provides data-driven insights, analyzes user data to identify emerging trends, collaborates with Alex to align market insights with the product vision, and works with Jamie to integrate market trends into the design process.

\vspace{0.2cm}

\noindent\textbf{Product Designer (PD) - Jamie}

\noindent\textit{Role:} Responsible for designing the product, ensuring its usability and aesthetics.

\noindent\textit{Responsibilities:} Jamie creates user-friendly and visually appealing designs, incorporates market trends and customer preferences, collaborates with Sam, and works with Taylor to align product design with sales strategies.

\vspace{0.2cm}

\noindent\textbf{Sales Manager (SM) - Taylor}

\noindent\textit{Role:} Develops sales strategies and manages customer relationships.

\noindent\textit{Responsibilities:} Taylor provides insights on customer preferences, sales potential, and go-to-market strategies, collaborates with Alex to ensure alignment with the product vision, works with Jamie to meet customer expectations, and gathers feedback to refine the product.

\subsection{Analysis of Agent Interactions and Decision-Making Processes}

This section explores how agents interact and make decisions during product development, covering key aspects such as:

\begin{itemize}
\item \textbf{Communication Protocol:}
\begin{itemize}
\item The PM (Alex) oversees interactions, ensuring each agent contributes based on their expertise.
\item Control returns to the PM after each agent's response, who then assesses the task and assigns the next one.
\item The structured format ensures efficient information flow and collaborative decision-making.
\end{itemize}

\item \textbf{Iterative Development:}
\begin{itemize}
    \item During the Build phase, Jamie develops the MVP, focusing on core features.
    \item Sam and Taylor provide input to ensure the MVP meets market demands and sales potential.
    \item Feedback is collected and analyzed to guide subsequent iterations.
\end{itemize}

\item \textbf{Decision-Making:}
\begin{itemize}
    \item Decisions are based on quantitative data and qualitative feedback.
    \item Alex synthesizes insights from all agents to decide whether to persevere or pivot.
    \item Documentation of each phase ensures transparency and informed decision-making.
\end{itemize}

\item \textbf{Conflict Resolution:}
\begin{itemize}
    \item Differences in opinions are addressed through structured discussions.
    \item The PM ensures all perspectives are considered, selecting the best course of action.
\end{itemize}

\end{itemize}

\section{Evaluation}

Comparing SynergyMAS \footnote{Code available at \url{https://github.com/feilaz/SynergyMAS-Evaluation}} directly with traditional systems like ChatGPT-4o, ChatGPT-4o with Chain of Thought (CoT), and ChatGPT-4o with Tree of Thoughts (ToT) is challenging due to their design differences \citep{wei2023chainofthoughtpromptingelicitsreasoning, yao2023treethoughtsdeliberateproblem, guo2024largelanguagemodelbased}. SynergyMAS excels in handling longer conversations and collaborative problem-solving, while the other models vary in their ability to maintain context and provide depth over extended interactions. To gain valuable insights, a controlled test scenario evaluated each system's performance in analyzing a Smart Home Energy Management System, a critical component in the Lean Startup methodology. The quality of each response was assessed to draw conclusions from the results.

\subsection{Comparative Analysis with Existing Multi-Agent Systems}

Structural Analysis: ChatGPT-4o provides a clear, concise structure but lacks detail in areas like energy savings variability. CoT offers a more comprehensive analysis but with some repetition. ToT enhances structure and provides future-oriented thinking with clear next steps. SynergyMAS balances structure and detail, offering specific data points and calculations, though with some inconsistencies in structuring.

Content Analysis: ChatGPT-4o focuses on key insights and next steps but lacks depth in areas like energy savings variability. CoT provides detailed analysis of specific data points, though with less emphasis on competitive analysis. ToT excels in a comprehensive, future-oriented analysis. SynergyMAS provides a thorough competitive analysis and justified priorities despite some redundancy in conclusions.

Analytical Depth: ChatGPT-4o offers a good overview but lacks depth in exploring causes of energy savings variability. CoT explores multiple factors in detail, providing AI improvement suggestions. ToT delivers the most in-depth analysis among single-model approaches, integrating future strategies. SynergyMAS offers the deepest analysis, exploring factors from multiple perspectives and providing a comprehensive strategy.

Unique Contributions: ChatGPT-4o emphasizes climate advantages and AI personalization. CoT analyzes energy savings variability with precise suggestions. ToT integrates a comprehensive, future-oriented perspective. SynergyMAS provides the most exhaustive competitive position analysis and detailed strategies for gamification, emphasizing user education for AI adoption.

\subsection{Discussion of Strengths and Limitations}
SynergyMAS excels in multi-perspective analyses with transparent reasoning processes. Its multi-agent design enhances collaborative problem-solving, effectively combining specialized knowledge \citep{wang2024sibylsimpleeffectiveagent}. The iterative improvement process and the ability to retrieve external information give SynergyMAS an advantage in complex scenarios. While ToT offers a highly structured, in-depth approach with a strong future-oriented perspective, SynergyMAS's ability to integrate diverse viewpoints and provide iterative improvements remains superior in complex scenarios. However, SynergyMAS could reduce redundancy in agent responses and improve analysis techniques. Its multi-agent nature may introduce variability \citep{chen2024commcollaborativemultiagentmultireasoningpath}, raising efficiency concerns compared to single-model approaches like ChatGPT-4o, CoT, and ToT.

\section{Future Work}

Scalability is vital for SynergyMAS's future. As tasks grow in complexity, the system must manage more agents and handle longer, more intricate conversations. Future efforts will focus on optimizing the architecture to meet these challenges while staying goal-oriented. Potential enhancements include refining the hierarchical structure or adopting more flexible nested designs \citep{han2024llmmultiagentsystemschallenges}, alongside advanced memory management techniques \citep{xie2024weknowragadaptiveapproachretrievalaugmented} and improved information synthesis across threads, making the framework more adaptable to evolving tasks \citep{wang2024megaagentpracticalframeworkautonomous}. These advancements will equip SynergyMAS to tackle increasingly complex problems while maintaining coherence and efficiency.

While SynergyMAS has been demonstrated with specific use cases, such as a Smart Home Energy Management System, future work will also involve testing the framework on open-source models like LLaMA, Gemma, and PHI-3. This will help assess its adaptability and performance across a broader range of LLM architectures. Additionally, the framework's principles can be adapted to various other domains, such as healthcare for collaborative diagnostics and finance for market analysis. Similar systems have shown promise in software development for project management \citep{cinkusz2024communicative, 10.1145/3691620.3695336}, and SynergyMAS could also support personalized learning in education. With the rapid emergence of new LLMs, expanding SynergyMAS to integrate with a wider variety of models will further enhance its applicability and effectiveness across diverse tasks and domains.

\section{Conclusion} 
 This paper has presented SynergyMAS, a multi-agent system designed to enhance collaborative problem-solving by integrating logical reasoning, RAG-based knowledge management, and Theory of Mind capabilities. The system's performance was evaluated in the context of developing a Smart Home Energy Management System, demonstrating significant improvements in analysis depth, data-driven insights, and actionable recommendations. SynergyMAS offers an ordered and iterative approach, using specialized agents to provide exhaustive and well-justified responses. 

 The development and evaluation of SynergyMAS contribute to the field of multi-agent LLM research by showcasing the potential benefits of integrating advanced reasoning and collaboration capabilities. The system's proficiency in handling complex tasks suggests that similar approaches could be beneficial in other domains requiring refined problem-solving and decision-making. By emphasizing the importance of scalability and versatility, this work lays the groundwork for future research and development in multi-agent systems, ultimately aiming to create more intelligent and effective AI solutions for various applications.

\section*{Acknowledgments}
We would like to acknowledge that the work reported in this paper has been supported in part by the Polish National Science Centre, Poland (Chist-Era IV) under grant 2022/04/Y/ST6/00001
\bibliography{acl_latex}

\end{document}